\documentclass[12pt,a4paper]{iopart}

\usepackage{iopams}							% packets for math
\usepackage{setstack}						% for e.g. subscripts in summation
\usepackage[english]{babel}			% encoding and language
\usepackage{graphicx}						% packets for pictures
\usepackage{acronym}						% acronyms
\usepackage{hyperref}						% links in pdf files
\usepackage{cite}								% citation
\usepackage[dvipsnames]{xcolor}

\newcommand{\ii}{\rmi}										% imaginary unit
\newcommand{\jj}{{\rm j}}									% imaginary unit
\newcommand{\kk}{{\rm k}}									% imaginary unit
\newcommand{\ee}{\rme}										% exponential function
\renewcommand{\Re}{{\rm Re}\,}						% real part
\renewcommand{\Im}{{\rm Im}\,}						% imaginary part
\newcommand{\PT}{\texorpdfstring{$\mathcal{PT}$}{PT}}

\renewcommand{\vec}[1]{\ensuremath{{\bf #1}}}
\renewcommand{\mat}[1]{\ensuremath{{\bf #1}}}

\newcommand{\braketop}[3]{\ensuremath{{\left\langle#1\middle|#2\middle|#3\right\rangle}}}

\newcommand{\bpparam}[1]{\ensuremath{{{#1}^{\oplus}}}}
\newcommand{\bmparam}[1]{\ensuremath{{{#1}^{\ominus}}}}
\newcommand{\bep}{\ensuremath{{\mathrm{e}^\oplus}}}
\newcommand{\bem}{\ensuremath{{\mathrm{e}^\ominus}}}
\newcommand{\bbraketop}[3]{\ensuremath{\left\langle#1\middle|#2\middle|#3\right\rangle_{\mathbb{BC}}}}

\begin{document}

% Abbreviations
\acrodef{BEC}{Bose-Einstein condensate}
\acrodef{GPE}{Gross-Pitaevskii equation}
\acrodef{TDVP}{time-dependent variational principle}

% -----------------------------------------------------------------------------
% title
%\title[Bifurcations structure and EPs \ldots]{
%	Bifurcation structure and exceptional points of the
%	analytically continued Gross-Pitaevskii equation for
%	dipolar Bose-Einstein condensates in an external
%	\texorpdfstring{\PT}{PT}-symmetric trap}
\title[Bifurcations and EPs in a PT-symmetric dipolar BEC]{Bifurcations and
	exceptional points in a \texorpdfstring{\PT}{PT}-symmetric 
	dipolar Bose-Einstein condensate}
\date{}
\author{Robin Gut\"ohrlein, Holger Cartarius, J\"org Main and G\"unter Wunner}
\address{1. Institut f\"ur Theoretische Physik,
  Universit\"at Stuttgart, 70550 Stuttgart, Germany}
\ead{Robin.Gutoehrlein@itp1.uni-stuttgart.de}

% -----------------------------------------------------------------------------
\begin{abstract}
	We investigate the bifurcation structure of stationary states in a dipolar
	Bose-Einstein condensate located in an external \PT-symmetric potential. The
	imaginary part of this external potential allows for the effective
	description of in- and out-coupling of particles.  To unveil the complete
	bifurcation structure and the properties of the exceptional points we perform
	an analytical continuation of the Gross-Pitaevskii equation, which is used to
	describe the system.  We use an elegant and numerically
	efficient method for the analytical continuation of the
	Gross-Pitaevskii equation with dipolar interactions containing
	bicomplex numbers. The Bose-Einstein condensate with dipole interaction shows
	a much richer bifurcation scenario than a condensate without long-range
	interactions. The inclusion of analytically continued states can also explain
	property changes of bifurcation points which were hidden before, and allows
	for the examination of the properties of the exceptional points associated
	with the branch points. With the new analytically continued
	states we are able to prove the existence of an exceptional point of fifth
	order.
\end{abstract}

\pacs{03.75.Kk, 11.30.Er, 03.65.Ge}
% 03.75.Kk		-	Bose-Einstein condensates dynamic properties
% 11.30.Er		-	Parity symmetry
% 03.65.Ge		-	Bound states quantum mechanics

% -----------------------------------------------------------------------------
\section{Introduction}
\label{sec:introduction}

Recently \PT-symmetric systems have gained much attention. Such systems do feature a
special class of non-Hermitian Hamiltonians which exhibit special properties
such as a real eigenvalue spectrum \cite{Bender1999a}. An operator is considered
to be \PT-symmetric if it commutes with the \PT-operator,
\begin{eqnarray}
	[\mathcal{PT}, H] = 0,
\end{eqnarray}
where $\mathcal{P}$ is the parity operator ($\hat x \rightarrow -\hat x$ and
$\hat p \rightarrow -\hat p$), and $\mathcal{T}$ is the time-reversal operator
which changes $\hat p \rightarrow - \hat p$ and $\ii \rightarrow -\ii$. For a
system of which the Hamiltonian can be written as
\begin{eqnarray}
	H = -\Delta + V(x),
\end{eqnarray}
\PT-symmetry implies the condition
\begin{eqnarray}
	\label{eq:ptpot}
	V(x) = V^*(-x)
\end{eqnarray}
for the (complex) potential. In such a system the imaginary part of the
potential represents the in- and out-coupling of particles into or from an
external reservoir. 

Quantum systems fulfilling this property have been studied in, e.g.,
\cite{qm1,qm2,Mehri,Bender1999a}. However, the concept of \PT-symmetry is not¬
restricted to quantum mechanics. The first experimental realization of
\PT-symmetric systems was indeed achieved in optical wave guides where the
effects of \PT-symmetry and \PT-symmetry breaking were observed
\cite{Rueter10}.  These first breakthroughs have increased the research effort
put into this field \cite{PhysRevA.88.053817, Deffner2015, Albeverio2015,
Mostafazadeh2013b}.  \PT-symmetric systems have also been studied in microwave
cavities \cite{Bittner2012a}, electronic devices
\cite{Schindler2011a,Schindler2012a}, and in optical \cite{0305-4470-38-9-L03,
opticPT1, ramezani10, musslimani08a, optic1, optic2, Makris2010a, makris08,
Chong2011, Peng2014} systems.  Also in quantum mechanics the stationary
Schr\"odinger equation was solved for scattering solutions \cite{qm2} and bound
states \cite{Mehri}. The characteristic \PT-symmetric properties are still found
when a many-particle description is used \cite{Dast2014}. In
\cite{Guthrlein2015} a \PT-symmetric system was embedded as a subsystem into an
hermitian system, showing that the subsystem retained its \PT-symmetric
properties.

In the first realizations of Bose-Einstein condensates
\cite{And95a,Bra95a,Dav95a} atoms without long-range interactions were used.
Here the dominating interactions can be described by s-wave scattering. However,
when other atoms with a large dipolar moment were investigated
\cite{Griesmaier2005,Santos2000,PhysRevA.77.061601, PhysRevLett.107.190401} new
effects emerged. Depending on the strength of the dipole-dipole interaction a
novel behaviour can be introduced into the condensates, such as an anisotropic
collapse of $^{52}$Cr atoms \cite{PhysRevLett.101.080401}.

Bose-Einstein condensates described by the Gross-Pitaevskii equation can be
placed in a \PT-symmetric potential. While new effects of
condensates with long-range dipole-dipole interactions have been found
\cite{Raghavan1999, Zaman2009, Xiong2009} when placed in a double-well trap,
completely new structures arise if additional gain and loss terms are
introduced. 

Different processes for the realisation of loss and gain in a condensate were
proposed. For the loss of particles the use of a focused electron beam
\cite{Gericke2008} at the loss site was examined. The gain might be realized by
feeding the condensate from a second condensate \cite{Robins2008} in a Raman
super-radiance-like pumping \cite{Schneble2004, Yoshikawa2004}. Also the coupling
to additional potential wells with a higher or lower potential base could be
used \cite{Kreibich2014}.

Since the \PT-symmetry of the
Gross-Pitaevskii equation depends on the \PT-symmetry of the wave function,
effects which change the geometry of the wave function can lead to additional
phenomena. Dipolar Bose-Einstein condensates exhibit such effects, e.g.\
structured ground states have been found \cite{PhysRevLett.98.030406}.
Therefore one would expect that the combination of dipolar Bose-Einstein
condensates with a \PT-symmetric trap will lead to a new behaviour. In
\cite{Fortanier2014} a Bose-Einstein condensate with long-range dipole-dipole
interaction in a \PT-symmetric double-well potential was examined.  This
condensate shows a richer, much more elaborate bifurcation scenario with more
states involved than in the case of a condensate with only short-range
interactions. Some of these bifurcations include up to five states,
and therefore allow for the possibile existence of higher-order exceptional
points, that is points in the parameter space where the eigenvalues and
eigenfunctions coalesce \cite{Kato66, Holger2007, Heiss2008, Graefe08a , Dem12,
Heiss2012, Heiss13a}.

An analytical continuation of the Gross-Pitaevskii equation provides the
mathematical tool to examine bifurcations and exceptional points in detail. An
encircling of exceptional points in complex parameter space can reveal, through
the exchange behaviour of the participating states, the order of the exceptional
point \cite{Guthrlein2013}. Also additional states and bifurcations which only
exist in the analytically continued space are revealed.  We can apply this
method to the system investigated in \cite{Fortanier2014} where bifurcations
with up to five states have been observed. 

In this paper we will examine this bifurcation scenario in much more detail
using an analytically continued Gross-Pitaevskii equation, therefore allowing us
to find additional (mathematical) states which compose the bifurcations. This
mathematical tool also allows us to expose the exchange behaviour of the
exceptional points associated with the bifurcations. In \sref{sec:gpe}
we will first give a short recapitulation of the Gross-Pitaevskii equation and
the ansatz of the wave function. In \sref{sec:ac} we will give an introduction
to the analytical continuation with bicomplex numbers and the representation in
an idempotent basis. Finally in \sref{sec:res} we will present new features in
the bifurcation scenario and examine the properties of the associated
exceptional points.

\section{Gross-Pitaevskii equation and the time-dependent variational principle}
\label{sec:gpe}

In this section we recapitulate the theoretical description of a Bose-Einstein
condensate using an ansatz of coupled Gaussians for the wave function. We use
the following parameterization of the Gross-Pitaevskii equation
\begin{equation}
	\label{eq:gpe}
	\left[
		- \Delta
		+ c_{\rm sc}  \left| \psi(\vec x, t) \right|^2
		+ V_{\rm d}
		+ V_{\rm ext}
	\right] \psi(\vec x, t)
	= \ii \frac{{\rm d}}{{\rm d} t} \psi(\vec x, t)
\end{equation}
with the scattering length $c_{\rm sc}$. The long-range dipole
interaction is described by the dipole potential
\begin{eqnarray}
	V_{\rm d} = c_{\rm d} \int {\rm d}^3 x' \frac{1-3\cos^2 \theta}{|\vec x - \vec
	x'|^3} \left| \psi(\vec {x'}, t) \right|^2
\end{eqnarray}
with the dipole strength $c_{\rm d} = 1$ and the angle $\theta$ between $\vec x
- \vec x'$ and the dipole alignment. A more complete overview of the
theoretical description and experimental realisation can be found in
\cite{Lahaye2009}.
We describe the external double-well
potential by two Gaussian beams \cite{Fortanier2014}
\begin{eqnarray}
	V_{\rm ext} &= \sum_{i=1}^{2} c_{{\rm g}, i}
		\exp\left( - (\vec x^T \mat A_{{\rm g},i} \vec x + \vec x^T \vec q_{{\rm g},i}
		+ \gamma_{{\rm g},i}) \right),
\end{eqnarray}
where $c_{{\rm g},i}$ defines the amplitude, $\mat A$, $\vec q$ and $\gamma$
define the shape and location of the Gaussian beams. The two wells are located
at $\vec q = ( \pm x_{\rm pos}, 0, 0 )$ symmetrically with respect to the origin
and with the amplitude
\begin{eqnarray}
 c_{{\rm g},1} = V_0 + \ii \Gamma
 ~{\rm and}~
 c_{{\rm g},2} = V_0 - \ii \Gamma,
\end{eqnarray}
and therefore fullfilling the requirement \eref{eq:ptpot} for a \PT-symmetric
potential. The parameter $V_0$ represents the strength of the real potential,
and $\Gamma$ determines the strength of the in- and outcoupling of particles
from an external reservoir. We are especially interested in how a change of the
in- and outcoupling strength $\Gamma$ effects the bifurcation behaviour.

To obtain the equations of motion we apply the time-dependent variational
principle by McLachlan \cite{McLachlan1964a} to \eref{eq:gpe}. To
fullfil the GPE as best as possible the difference betweeen the two sides of
the equation
\begin{eqnarray}
	I = || \ii\psi - H \underbrace{\frac{\partial}{\partial t} \psi}_{=: \phi} ||^2
\end{eqnarray}
is minimized. For a given time $t$ the wave function $\psi$ is pinned and the
time-derivative $\phi$ is varied. Hence, if $I$ is a minimum for small $\delta\phi$
the change of $\delta I$ must vanish. We make an ansatz of coupled
Gaussians for the wave function
\begin{eqnarray}
	\label{eq:ansatz}
	\psi(\vec x, t)
	= \sum_{k=1}^N g^k
	= \sum_{k=1}^N \exp\left(-\left(
				\vec x^T \mat A_k \vec x
			+ \vec p_k^T \vec x
			+ \gamma_k
		\right)\right)
\end{eqnarray}
where the complex symmetric matrix $\mat
A$ defines the shape of the Gaussian, the complex momentum vector is given by
$\vec p$ and the complex number $\gamma$ defines the amplitude and phase of the
Gaussian. This ansatz does not allow for arbitrary matrices
$\mat A$, e.g. matrices which contain a negative gaussian widths are not
normalizable and are not accepted as solutions.

Because our trap exhibits a strong confinement in the $y$ direction
\cite{Fortanier2014} we can introduce further restrictions to the ansatz of our
wave-function. We do not consider all rotational and translational degrees of freedom in
the direction of the confinement ($A_{xy} = A_{yz} = p_{y} = 0$). With the help
of the time-dependent variational principle one can obtain the time derivatives
of the Gaussian parameters $\dot\mat A_k$, $\dot\vec p_k$ and $\dot \gamma_k$
\cite{Fortanier2014}. A numerical root search can be performed to
obtain the stationary states. It was shown \cite{Rau10c} that the
variational principle, while being numerically cheap, provides results with
extremely high quality. For all calculations in this paper, an ansatz
of $N=2$ coupled Gaussians was chosen, i.e.\ the wave function is approximated
in each well by one Gaussian function (but they are not restricted to a specific
location).

\section{Analytical continuation}
\label{sec:ac}

The non-analyticity of the Gross-Pitaevskii equation \eref{eq:gpe},
due to the square modulus of the wave-function, prevents us
from examining the exchange behaviour of exceptional points. Also since the
equation is not analytical the number of states can change from one side of the
bifurcation to the other \cite{Fortanier2014}. It has been shown for simpler
systems without long-range interactions that with an analytical continuation
\cite{Guthrlein2013,Dast2013} the whole bifurcation scenario of the equation is
revealed and the complete relation between the different stationary states can
be observed. But since the Gross-Piaevskii equation is non-analytical
and already in the complex domain, the analytical continuation has to go beyond
the complex domain. One way to perform the analytical continuation is
to separate all complex equations and complex parameters into twice the number
of real equations with twice the number of real parameters. Then these equations
can be analytically continued. An equivalent alternative is the introduction of
bicomplex numbers.  In this approach the complex numbers $\mathbb{C}$ are
replaced by bicomplex numbers $\mathbb{BC}$ \cite{LUNA-ELIZARRARAS2012}.  Both
methods are mathematically completely equivalent, however, the second approach
is much more elegant since it provides huge numerical benefits as we will see in
the following discussion. Note that there are different hypercomplex
numbers. But for the description we have in mind only bicomplex numbers have the
correct properties. Furthermore this choice ensures that while extending the
original model, all its properties and states are contained and no imformation
is lost. In other scenarios different hypercomplex numbers are used and
appropriate, e.g. quaternions and co-quaternions are discussed in
\cite{Brody2011a, brody2011b}.

\subsection{Bicomplex numbers}

We first recapitulate some basic properties of bicomplex numbers
\cite{LUNA-ELIZARRARAS2012}. In contrast to complex numbers bicomplex numbers
have three imaginary units with the following properties
\begin{equation}
  \kk = \ii\jj ~,~~ \ii^2 = \jj^2 = -1 ~,~~ \kk^2 = 1.
\end{equation}
To analytically continue a complex equation we split the equation into its real
and imaginary part
\begin{equation}
	z =       
	  \underbrace{
	          \underbrace{z_{\rm r}}_{\in\mathbb{R}}
	    + \ii \underbrace{z_{\rm i}}_{\in\mathbb{R}}
		}_{\in \mathbb{C}}.
\end{equation}
These real and imaginary parts can be extended to complex numbers with the
imaginary unit $\jj$. Then $z$ becomes a bicomplex number
\begin{equation}
	\label{eq:binumber}
	z =  
		\underbrace{
			     \underbrace{(
										\underbrace{z_0}_{\in\mathbb{R}}
							+ \jj \underbrace{z_1}_{\in\mathbb{R}}
						)}_{\in \mathbb{C}_\jj}
	   + \ii \underbrace{(
										\underbrace{z_2}_{\in\mathbb{R}}
							+ \jj \underbrace{z_3}_{\in\mathbb{R}}
						)}_{\in \mathbb{C}_\jj}
		}_{\in \mathbb{BC}}
	  = z_0 + \jj z_1 + \ii z_2 + \kk z_3.
\end{equation}
The complex numbers $\mathbb{C}_\jj$ with the imaginary unit $\jj$ are
isomorphous to the complex numbers $\mathbb{C}_\ii$ with the imaginary unit
$\ii$.

\subsection{Representation of bicomplex numbers in the idempotent basis}

There exists a representation of bicomplex numbers in an idempotent basis. Let
us first consider the two idempotent elements
\begin{eqnarray}
	\bep = \frac{1+\kk}{2}
	~{\rm and}~
	\bem = \frac{1-\kk}{2},
\end{eqnarray}
which fullfil the relations
\begin{eqnarray}
	\bep^2 = \bep, ~\bem^2 = \bem ~{\rm and}~ \bep \bem = 0.
\end{eqnarray}
These properties allow us to decompose every bicomplex number of the form
\eref{eq:binumber} into
\begin{eqnarray}
	z &= \left( (z_0 + z_3) + (z_2 - z_1) \ii \right) \bep
		+ \left( (z_0 - z_3) + (z_2 + z_1) \ii \right) \bem \nonumber \\
		&= \bpparam{z} \bep + \bmparam{z} \bem.
\end{eqnarray}
The coefficients $\bpparam{z}$ and $\bmparam{z}$ can be chosen to be either
elements of $\mathbb{C}_\ii$ or $\mathbb{C}_\jj$. In our calculations we choose
the $\mathbb{C}_\ii$ representation. We can now consider the implication the
idempotent properties have on the basic arithmetic operations. One finds that
\cite{LUNA-ELIZARRARAS2012}
\numparts
\begin{eqnarray}
	a \pm b &= 
			(\bpparam{a}\pm\bpparam{b}) \bep
		+	(\bmparam{a}\pm\bmparam{b}) \bem,
	\\
	a \cdot b &= 
			(\bpparam{a}\cdot\bpparam{b}) \bep
		+	(\bmparam{a}\cdot\bmparam{b}) \bem,
	\\
	a / b &= 
			(\bpparam{a}/\bpparam{b}) \bep
		+	(\bmparam{a}/\bmparam{b}) \bem,
\end{eqnarray}
\endnumparts
viz.\ the operations can be performed independently for the plus and minus
components in the complex subspaces.  These properties allow us to use a very
efficient numerical implementation (which uses already existing algorithms).  In
particular for terms which contain the dipolar potential a highly optimized
algorithm for complex numbers is available. If the method without
bicomplex numbers would have been used other integrals had to be solved,
for which no algorithms exist. We will see in the next section that
special care must be taken if a calculation involves the complex conjugate.

\subsection{Complex conjugation}

There exist different kinds of complex conjugations for bicomplex numbers.
In \cite{Bagchi2015} possible conjugations and new \PT-symmetries
are discussed. However, for our application only one is relevant, since our
goal is that the extended model contains all states and properties of the
original model. In order to determine which one must be used we first
consider the complex conjugation which occurs in \eref{eq:gpe} without the
bicomplex continuation. We start with the complex conjugation of the real and
imaginary part of the complex number
\begin{eqnarray}
 	z^* &= z_{\rm r} - \ii z_{\rm i}.
\end{eqnarray}
Now the analytical continuation is applied by replacing the real and imaginary
parts with complex numbers with the imaginary unit $\jj$
\begin{eqnarray}
	z^* &= (z_0 + \jj z_1) - \ii (z_2 + \jj z_3) = z_0 + \jj z_1 - \ii z_2 - \kk z_3.
\end{eqnarray}
This complex conjugation changes the sign of all $\ii$ and $\kk$ components.
Using this complex conjungation we can write the modulus squared of the
wave function in \eref{eq:gpe} as
\begin{eqnarray}
	|\psi|^2 = \left<\psi \middle| \psi \right>_\mathbb{BC} = \psi^* \psi.
\end{eqnarray}
This complex conjugation also must be used in inner products. Note that in the
representation of the idempotent basis the complex conjugate reads
\begin{eqnarray}
	\label{eq:cc}
	z^* = \overline\bmparam{z} \bep + \overline\bpparam{z} \bem,
\end{eqnarray}
where the bar denotes the complex conjugation with $\bpparam{z}, \bmparam{z}
\in \mathbb{C}_\ii$.

\subsection{Decomposition of the GPE using the idempotent basis}

With the properties shown so far we can examine the analytical continuation
of the Gross-Pitaevskii equation. A linear set of equations, such as the one
obtained by the time-dependent variational principle \cite{FortanierDoc}, can be
decomposed into two linear systems of equations using the idempotent basis
\numparts
\begin{eqnarray}
	\bpparam{\mat K} \bpparam{\vec v} &= \bpparam{\vec r}, \\
	\bmparam{\mat K} \bmparam{\vec v} &= \bmparam{\vec r}.
\end{eqnarray}
\endnumparts
The vectors $\bpparam{\vec r}$, $\bmparam{\vec r}$ and the matrices
$\bpparam{\mat K}$, $\bmparam{\mat K}$ contain both plus and minus coefficients
of the Gaussian parameters. The elements of the bicomplex matrices $\mat K$ and
vectors $\vec r$ are of the form \cite{FortanierDoc}
\begin{eqnarray}
	\bbraketop{g_l}{\alpha^n \beta^m V}{g_k}
\end{eqnarray}
with $\alpha,\beta = x,y,z$ and $n, m \in \mathbb{N}$, and the Gaussians $g_l,
g_k$. If the bicomplex conjugate described in \eref{eq:cc} is used, one can
decompose the expression into
\begin{eqnarray}
	\bbraketop{g_l}{\alpha^n \beta^m}{g_k}
	=
	\braketop{\bmparam{g_l}}{\alpha^n \beta^m \bpparam{V}}{\bpparam{g_k}} \bep
	+
	\braketop{\bpparam{g_l}}{\alpha^n \beta^m \bmparam{V}}{\bmparam{g_k}} \bem
	.
\end{eqnarray}
This representation allows us to use an existing algorithm for complex numbers
to calculate the necessary integrals. The algorithm just must be called twice
for the appropriate parameter sets. For elliptic integrals the Carlson algorithm
\cite{Carlson} can be used. However, the largest numerical improvement is
achieved by using the highly optimized algorithm presented in
\cite{FortanierDoc} for the numerical integration of the dipolar term.

Finally we examine how the control parameters, which were real numbers before
the analytical continuation was applied, can be represented in the idempotent
basis. Originally the scattering length $c_{\rm sc}$ was real. After the complex
continuation it assumes the form
\begin{eqnarray}
	c_{\rm sc} = c_{\rm sc}^0 + \jj c_{\rm sc}^1,
\end{eqnarray}
with $c_{\rm sc}^0, c_{\rm sc}^1 \in \mathbb{R}$. We can now represent this
parameter in the idempotent basis
\begin{eqnarray}
	c_{\rm sc} &= c_{\rm sc}^0 + \jj c_{\rm sc}^1
		= \underbrace{(c_{\rm sc}^0 - \ii c_{\rm sc}^1)}_{\bpparam{c_{\rm sc}}} \bep +
			 \underbrace{(c_{\rm sc}^0 + \ii c_{\rm sc}^1)}_{\bmparam{c_{\rm sc}}} \bem.
\end{eqnarray}
Since there are only two idependent real parameters the relation
$\overline{\bpparam{c}_{\rm sc}} = \bmparam{c}_{\rm sc}$ is fullfilled. The same
also applies for the in- and out-coupling parameter $\Gamma$. 

Another special parameter type exists. If the bicomplex number $z$ only
contains a real part and an imaginary part for the complex unit $\ii$, but not
for $\jj$ or $\kk$, the number can be decomposed in the idempotent basis in such
a way that the two coefficients are equal, i.e.\ $\bpparam{z} = \bmparam{z}$.

\section{Results}
\label{sec:res}

\subsection{Bifurcation scenario of a BEC without long-range interaction}

To compare the bifurcation scenario between a BEC with and without dipolar
interaction, a short recapitulation of the bifurcation scenario of a BEC
without long-range interactions is given in this section. This scenario has been
discussed in detail in previous publications \cite{Graefe12,
Cartarius12b,Dast13,Haag14}. The bifurcation scenario of such a \PT-symmetric
double well system can be described by the matrix equation
\begin{eqnarray}
	\label{eq:mm}
	\left(\begin{array}{cc}
		-g |\psi_1|^2 - \ii \gamma & v \\
		v & -g |\psi_2|^2 + \ii \gamma
	\end{array}\right)
	\left(\begin{array}{c}
		\psi_1 \\ \psi_2
	\end{array}\right)
	= \mu
	\left(\begin{array}{c}
		\psi_1 \\ \psi_2
	\end{array}\right)
\end{eqnarray}
with the coupling strength $v$, the nonlinearity $g$, and the in- and
outcoupling strength $\gamma$.

\begin{figure}
	\includegraphics[width=0.98\textwidth]{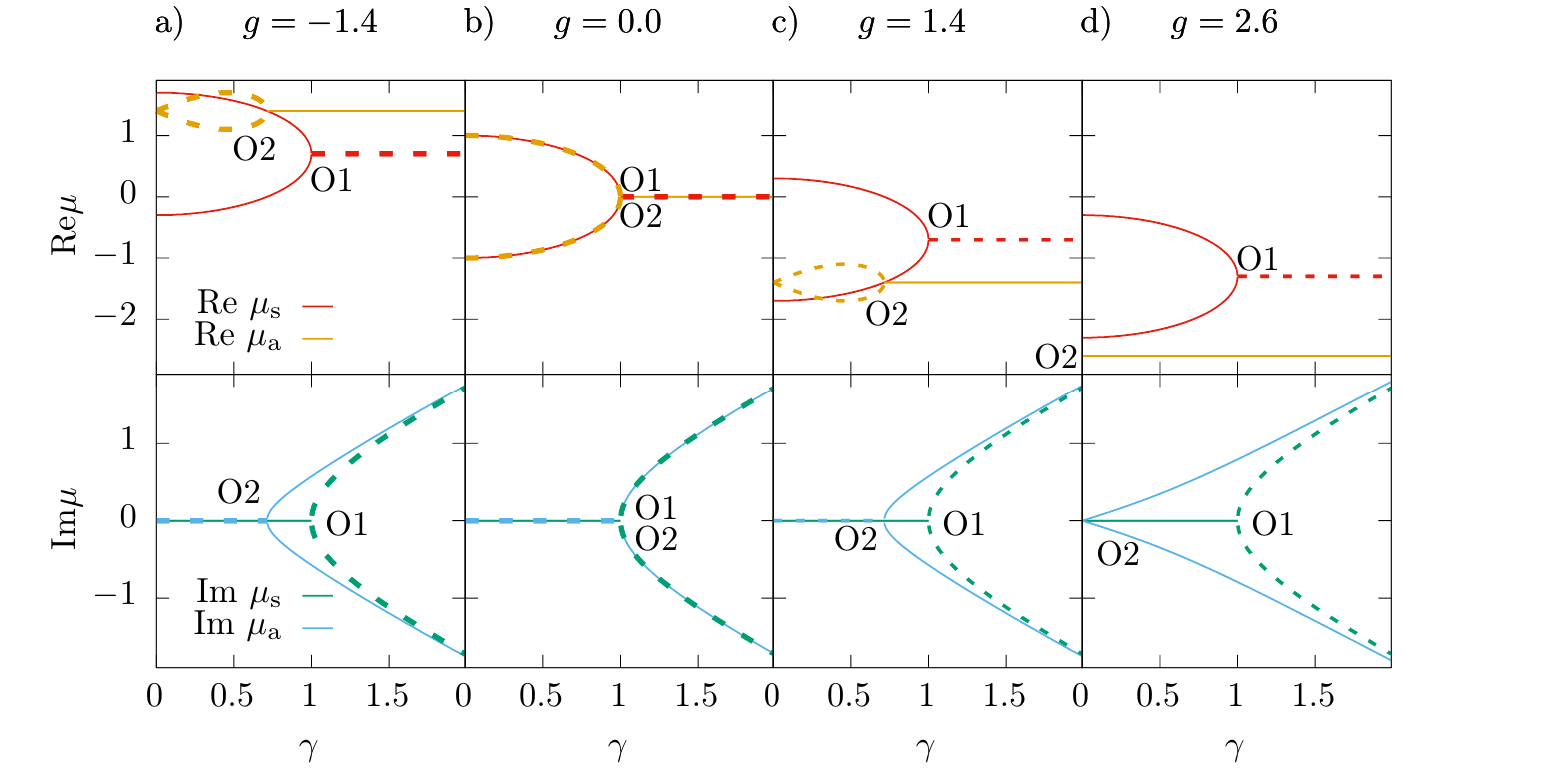}
	\caption{
		\label{fig:old}
		Bifurcation scenarios of a Bose-Einstein condensate without long-range
		interaction. The underlying equation \eref{eq:mm} was analytically continued
		using bicomplex numbers.  Solutions which exist in the complex plane are
		represented by solid lines.  Solutions which require the analytical
		continuation are shown as dashed lines. States which are \PT-symmetric are
		denoted by the subscript s, while states which break this symmetry are
		denoted by a.  }
\end{figure}

In the linear case $g = 0$ there exist two parameter ranges for the in- and
out-coupling strength $\gamma$. In the regime below a critical parameter
$\gamma_c$ there are two \PT-symmetric solutions, which merge in a tangent
bifurcation (O1 in \fref{fig:old}(b)) at the critical
parameter $\gamma_c$.  For values greater than this critical strength, there
exist two \PT-broken solutions with complex chemical potentials. The tangent
bifurcation in this scenario is a second-order exceptional point. One can
observe the pitchfork bifurcation (O2 in \fref{fig:old}) which occurs on the
upper \PT-symmetric branch for values of $-2 < g < 0$ (see \fref{fig:old}(a))
and switches to the lower \PT-symmetric branch for values of $0 < g < 2$ (see
\fref{fig:old}(c)). This bifurcation vanishes for values $g < -2$ or $g > 2$
(see \fref{fig:old}(d)).

We will see that a bifurcation similar to bifurcation O1 appears in a dipolar
Bose-Einstein condensate. The behaviour of the bifurcation O2, which also
occurs, will be altered in certain parameter regions. The change in behaviour is
also dependent on new emerging bifurcations.

\subsection{Results for a BEC with long-range dipole interactions}

The calculations for the analytically continued GPE \eref{eq:gpe} reveales the
complete bifurcation structure of the system. In addition to the known
stationary states which exist in the complex domain new states which are truly
bicomplex are found. With these new states all branches of the bifurcations are
present.  In addition it is possible to evaluate the exchange behaviour of the
exceptional points.

\begin{figure}
	\includegraphics[width=0.98\textwidth]{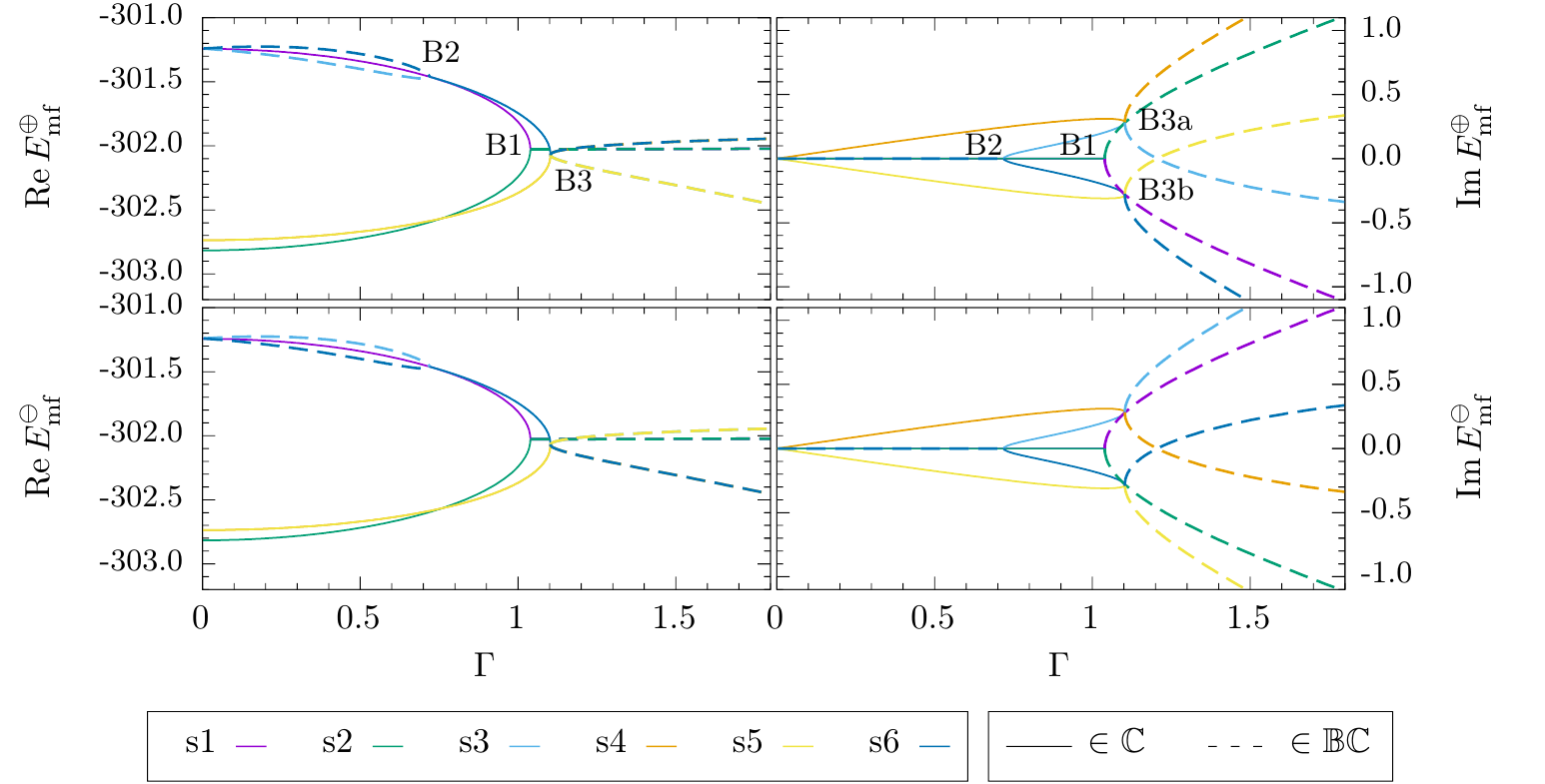}
	\caption{
		\label{fig:line0}
		Real and imaginary parts of the plus and minus component of the bicomplex
		mean-field energy in a dipolar Bose-Einstein condensate with scattering
		length $c_{\rm sc} = -0.9$. The mean-field energy is composed as $E_{\rm mf}
		= \bpparam{E_{\rm mf}} \bep + \bmparam{E_{\rm mf}} \bem$. States which
		exist without the analytical continuation into bicomplex values must obey
		$\bpparam{E_{\rm mf}} = \bmparam{E_{\rm mf}}$ and are shown as solid lines.
		States which exist only in the bicomplex domain are shown as dashed lines.
		The four bifurcations which occur between the states are marked with B1 to
		B4.  As discussed in \cite{Fortanier2014} states s1 and s2 obey \PT-symmetry
		from $\Gamma = 0$ up to the bifurcation B1. Beyond this point the states are
		\PT-broken.  }
\end{figure}

In \fref{fig:line0} eigenvalues for states with $c_{\rm sc} = -0.9$ are shown.
The states which are purely complex (solid lines) are already known from
\cite{Fortanier2014}. Since these states are complex the relation
\begin{equation}
	\bpparam{E_{\rm mf}} = \bmparam{E_{\rm mf}} \label{eq:rel_energy_complex}
\end{equation}
holds true. The real and imaginary parts of the mean-field energy
are
\begin{eqnarray}
	\Re E_{\rm mf} = \frac{1}{2}   \left( \bpparam{E_{\rm mf}} + \bmparam{E_{\rm mf}} \right)
	~~{\rm and}~~
	\Im E_{\rm mf} = \frac{\ii}{2} \left( \bpparam{E_{\rm mf}} - \bmparam{E_{\rm mf}} \right)
	.
\end{eqnarray}
The imaginary part of the mean-field energy exists due to the loss or gain of
particles in the system. For the states present in the original model the real
and imaginary parts of the energy are real numbers. For new analytically
continued states these numbers are now complex. The relation
\eref{eq:rel_energy_complex} no longer holds true for states living only in the
bicomplex space (dashed lines).  Since all branches of the
bifurcations are now present, there exist six states for the whole parameter
range of $\Gamma$.  The chemical potential exhibits the same qualitative
behaviour as the mean-field energy.  Since the qualitative
behaviour is the same in the plus and minus components only the plus component
of the mean-field energy is discussed in the following.

% Bifurcation B1

\begin{figure}
	\includegraphics[width=0.98\textwidth]{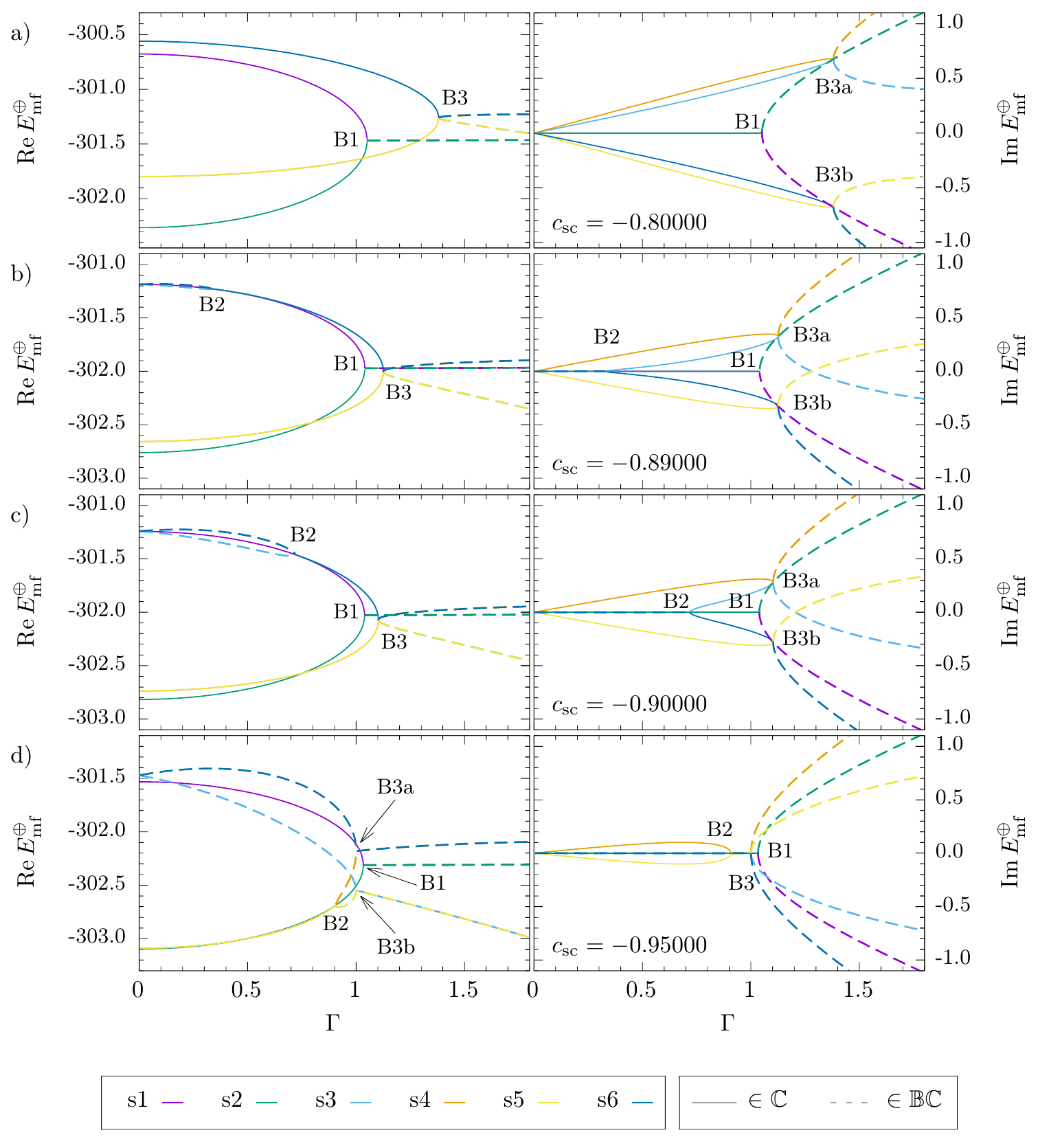}
	\caption{
		\label{fig:line1}
		Mean-field energies $\bpparam{E_{\rm mf}}$ for different values of $c_{\rm
		sc}$. The different bifurcation points are denoted by B1, B2, B3a and B3b.
		When the bifurcations B3a and B3b coincide they are marked by B3. The
		bifurcation B2 does not exist in (a).  }
\end{figure}

In \fref{fig:line1} the real and imaginary parts of the plus component of the
mean-field energy are shown for different values of the scattering strength. For
small values of $\Gamma$ the states s1 and s2 show the typical behaviour also
known from BECs without long-range interactions. They merge in a tangent
bifurcation B1. Up to this point the states also exist without the analytical
continuation. For larger $\Gamma$ they become bicomplex. This qualitative
behaviour is independent of the scattering length, only the critical point is
slightly shifted.

% Bifurcation B2
\begin{figure}
	\includegraphics[width=0.98\textwidth]{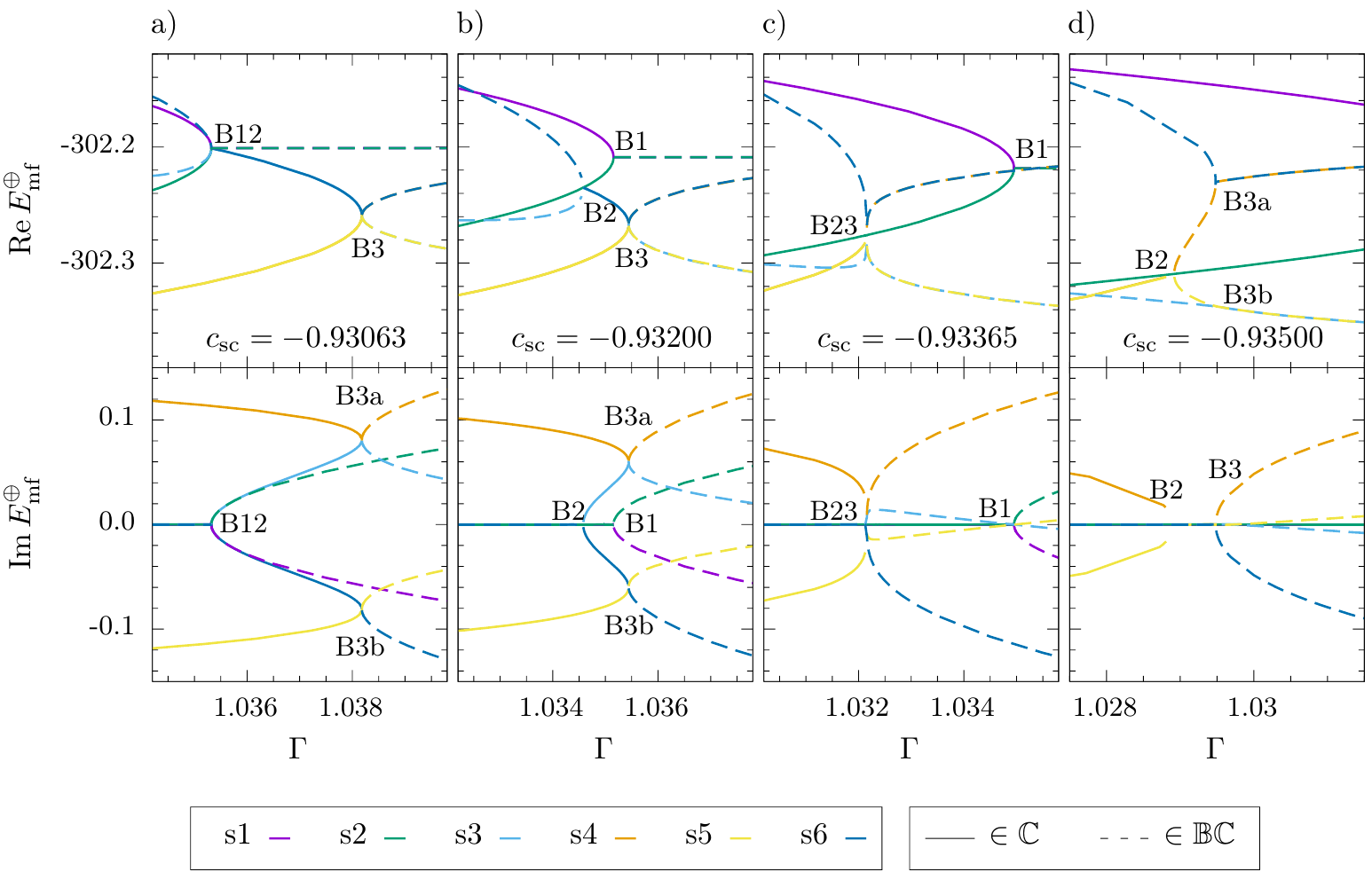}
	\caption{
		\label{fig:line2}
		For a scattering length of $c_{\rm sc,crit,1} = -0.93063$ the bifurcations
		and EP2s B1 and B2 merge. This can be seen in subfigure a) and the resulting
		bifurcation is marked as B12. In subfigure c) the EPs and bifurcations B2,
		B3a and B3b merge for $c_{\rm sc,crit,2} = -0.93365$ into a higher order
		exceptional point which is denoted by B23. Subfigure b) shows the
		bifurcation scenario for an intermediate scattering length of $c_{\rm sc} =
		-0.932$.  }
\end{figure}

The bifurcation B2 undergoes multiple behaviour changes for different scattering
lengths. Also this bifurcation does not always exist (see \fref{fig:line1}(a)).
For smaller scattering lengths the bifurcation appears at $\Gamma = 0$. This is
a pitchfork bifurcation between the states s1, s3, and s6. The bifurcation moves
along the state s1 to larger $\Gamma$ as the scattering length is decreased (see
\fref{fig:line1}(b) and (c)). At a critical scattering length $c_{\rm sc, crit,
1} = -0.93063$ the bifurcation merges with the bifurcation B1 (see
\fref{fig:line2}(a)). For even smaller scattering lengths the bifurcation moves
back to $\Gamma = 0$ along the state s2. The pitchfork bifurcation is now formed
between the states s2, s3, and s6 (see \fref{fig:line2}(b)). However, before the
bifurcation reaches $\Gamma = 0$ another critical scattering length is reached.
At $c_{\rm sc, crit, 2} = -0.93365$ the bifurcation B2 merges with the
bifurcations B3a and B3b (see \fref{fig:line2}(c)). At this point the behaviour
of the bifurcation is altered.  For larger scattering lengths at smaller
$\Gamma$ the participating states exist only for the bicomplex equations (see
\fref{fig:line2}(b)), but for larger $\Gamma$ the states exist in the complex
Gross-Pitaevskii equation (see \fref{fig:line2}(d)) without the bicomplex
extension. If the scattering length is smaller than $c_{\rm sc, crit, 2}$ this
behaviour is mirrored with respect to the $\Gamma$-axis. Until the bifurcation
vanishes for smaller scattering lengths at $\Gamma = 0$ it is composed of the
states s2, s4, and s5. If the bifurcation B2 is compared with the bifurcation O2
of the Bose-Einstein condensate without long-range interactions (see
\fref{fig:old}), the first change (the merger of B1 and B2) can also be
observed. However, the second change in behaviour is a new effect since the
bifurcations B3a and B3b do not exist without long-range interactions.

% Bifurcations B3a,B3b
We have seen that the merger of the bifurcations B2, B3a, and B3b changes the
behaviour of bifurcation B2. The properties of the tangent bifurcations B3a and
B3b are altered. For scattering lengths greater than $c_{\rm sc, crit, 2}$ the
bifurcations B3a and B3b are bicomplex for larger $\Gamma$ (i.e., the states
have components of the imaginary units $\jj$ and $\kk$). However, for $\Gamma$
below the critical value only the real component and the component with complex
unit $\ii$ is nonzero (see \fref{fig:line2}(b)). For smaller values of the
scattering lengths the states in both $\Gamma$ regions are bicomplex, i.e., they
only exist as solutions of the analytically continued Gross-Pitaevskii equation
(see \fref{fig:line2}(d)).

\begin{figure}
	\includegraphics[width=0.98\textwidth]{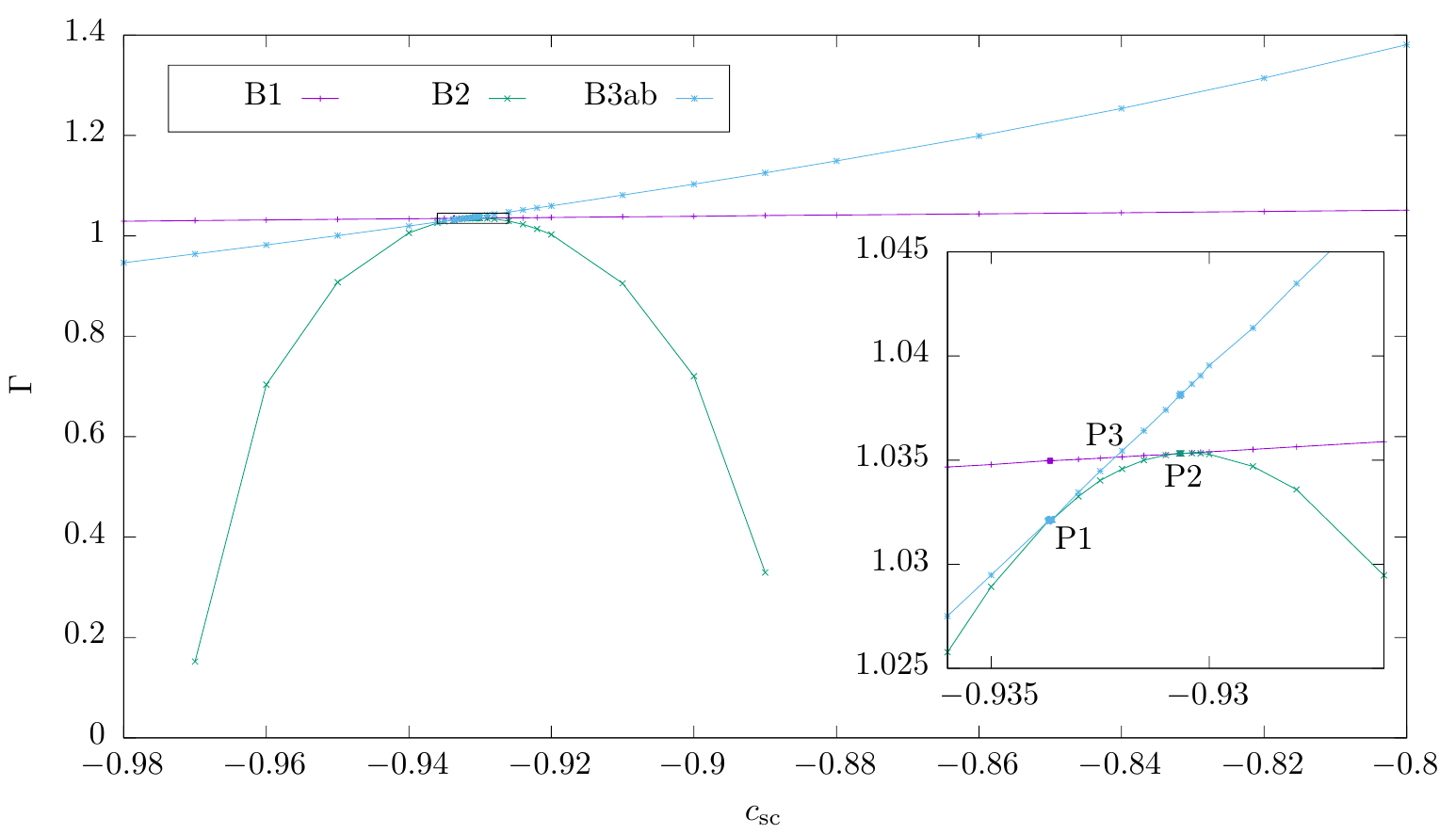}
	\caption{
		\label{fig:scgamma}
		Positions of the bifurcations and exceptional points in the $\Gamma$-$c_{\rm
		sc}$-parameter space. In the inset the merger of multiple bifurcations can
		be observed. Note that at the point P3 no merger occurs. There merely exist
		two bifurcations between different states at the same point in parameter
		space.  }
\end{figure}

% Parameter space
We have found that the critical scattering length parameters, which divide the
parameter region with different behaviours, are related with the merger of
multiple bifurcations.  In \fref{fig:scgamma} the parameter pairs of the
scattering length $c_{\rm sc}$ and $\Gamma$ are shown at which the different
bifurcations occur. One observes three points at which the parameters of two
different bifurcations become identical. Point P3 is not special, the states
which are involved in the two bifurcations (B1 und B3) have different
eigenvalues and wave functions.  Therefore just two independent bifurcations
occur at the same parameter pair.  By contrast at P2 the bifurcations B1 and B2
are joined into one bifurcation.  The bifurcation scenario is shown in
\fref{fig:line2}(a). Another bifurcation merger appears for the parameters at
point P1. The resulting bifurcation, which consits of B2, B3a, and B3b is shown
in \fref{fig:line2}(c).  These merger points are also of special interest
because they have the prerequesites necessary that exceptional points of higher
order can appear.

\subsection{Exchange behaviour of the states around the exceptional points}

We now examine which signatures of exceptional points can be
observed. In \cite{Dem12,Guthrlein2013} it was discussed that a complex
encircling of a higher order exceptional point does not have to exhibit an
exchange of all states which are involved in the exceptional point. Using
different parameters to encircle the EP can show different exchange behaviour.

\begin{figure}
	\includegraphics[width=0.98\textwidth]{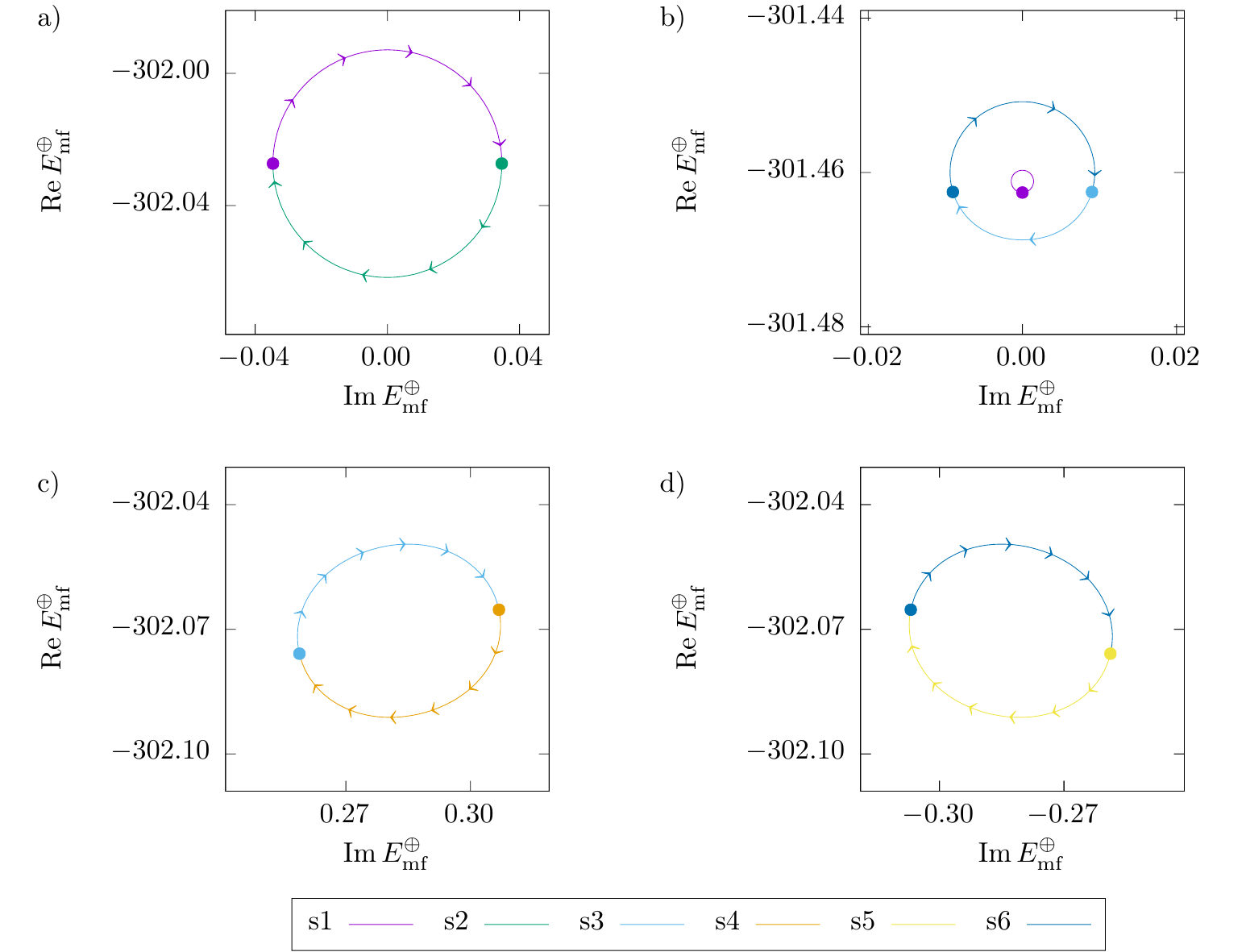}
	\caption{
		\label{fig:ep2}
		Characteristic exchange of states when an exceptional
		point is encircled in the complex parameter space. In this case the
		bifurcation B1  (subfigure (a), $\Gamma_{\rm center} = 1.03904$, $r = 10^{-3}$),
		bifurcation B2  (subfigure (b), $\Gamma_{\rm center} = 0.71967$, $r = 2\times10^{-3}$),
		bifurcation B3a (subfigure (c), $\Gamma_{\rm center} = 1.10296$, $r = 10^{-4}$) and
		bifurcation B3b (subfigure (d), $\Gamma_{\rm center} = 1.10296$, $r = 10^{-4}$)
		are encircled on the parameter path
		$\Gamma(x) = \Gamma_{\rm center} + r \ee^{\jj \phi}$ for $\phi \in [0,2\pi]$.
		All plots were calculated for a scattering length of $c_{\rm sc} = -0.9$.
	}
\end{figure}

The second-order exceptional points of the tangent bifurcations B1, B3a, and B3b
show the expected square root behaviour by exchanging each state with the other
(see \fref{fig:ep2}(a,c,d)) when the point is encircled in the complex parameter
space. On the other hand the third-order exceptional point at the bifurcation B2
only shows the exchange of two states. For the encircling in the complex
$\Gamma$ a cubic root exchange behaviour cannot be observed (see
\fref{fig:ep2}(b)).

\begin{figure}
	\includegraphics[width=0.98\textwidth]{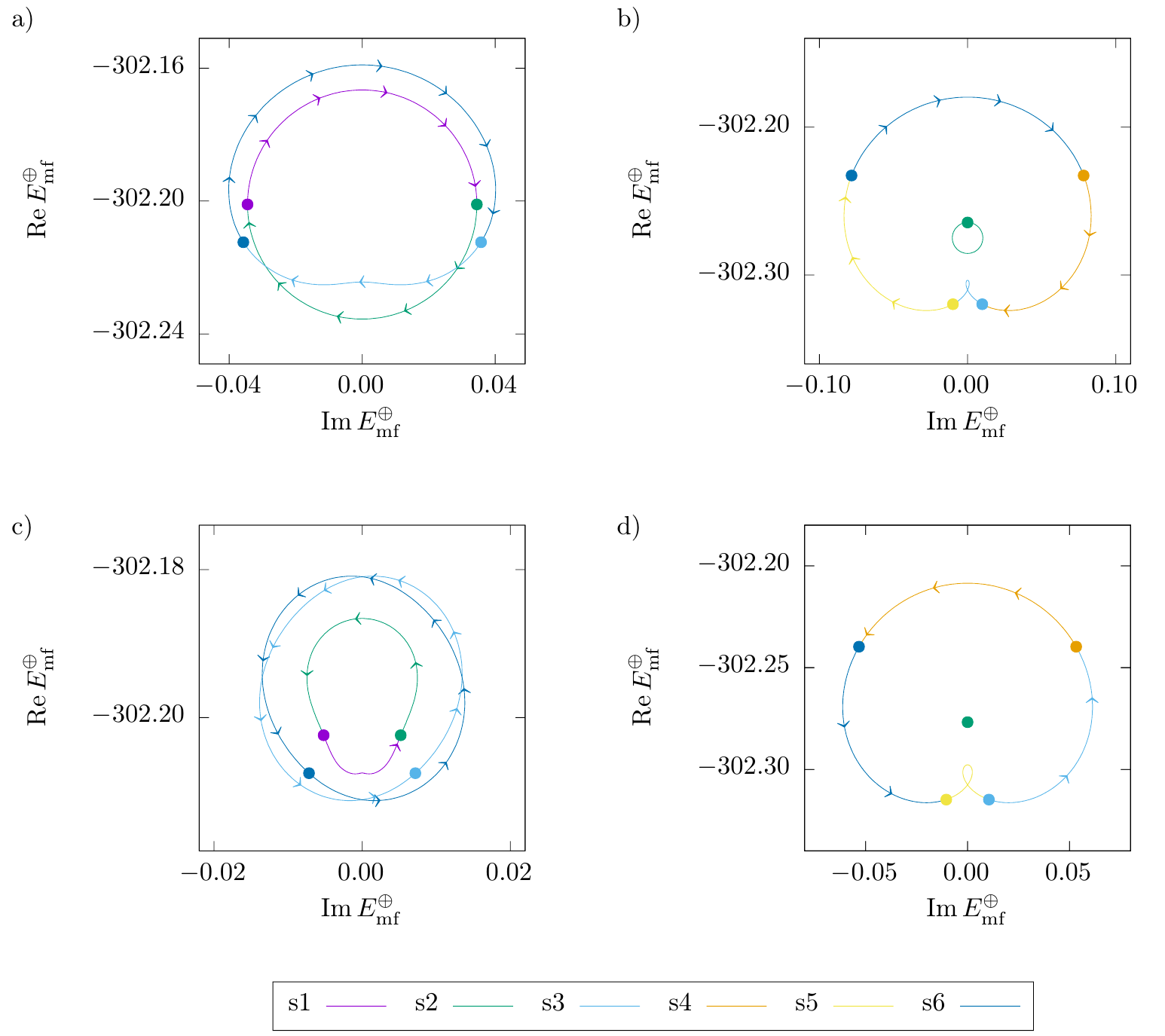}
	\caption{
		\label{fig:ep4ab}
		(a) and (c) show the exceptional point at the coalescence of bifurcation B1
		and B2 at $c_{\rm sc,crit,1} = -0.93063$ and $\Gamma = 1.03531$.  (b) and
		(d) show the exceptional point for the merger of the bifurcations B2, B3a
		and B3b at $c_{\rm sc,crit,2} = -0.93365$ and $\Gamma = 1.03215$. The
		exceptional points are encircled in the complex parameter space of $\Gamma$
		(subfigure (a) and (b)) with a radius of $r = 10^{-3}$. In (c) and (d) the
		exceptional points are encircled in the complex parameter space of $c_{\rm
		sc}$. The radius of the circle is $r = 4.65\times10^{-4}$ in (c) and $r =
		2\times10^{-4}$ in (d).  One exceptional point (subfigure (b) and (d))
		exhibits the full exchange behaviour of a fourth-order exceptional point, in
		which the original situation is not reached until the point is encircled
		four times. By contrast the other exceptional point in (a) and (c) has two
		separated groups of states which only exchange with each other, and
		therefore the starting condition is already reached again after two full
		orbits.  }
\end{figure}

In \fref{fig:ep4ab}(a) we show the exchange behaviour of the states involved in
P2 where the bifurcations B1 and B2 merge. When encircling the exceptional point
in a complex $\Gamma$-parameter plane an exchange within pairs of states is
found, however, these two exchanges are separated and no exchange between all
four states can be observed. Therefore it is unclear whether these are two
second-order exceptional points or one fourth-order exceptional point. Also by
encircling the critical point in the complex scattering length plane does not
change the qualitative behaviour, again only two states exchange.  To prove that
this must be an EP4, one must search for further complex perturbation
parameters.  However, since for an exceptional point of order $n$, all $n$
eigenvalues and eigenstates must coalesce \cite{Kato66} we examine the wave
functions of the participating states and they all coalesce at the critical
point (which means that for the ansatz of coupled Gaussians all Gaussian
parameters must be the same, which is indeed the case).

The same examination can be performed for the merger of the bifurcations B2, B3a,
and B3b (point P1). At this critical point five eigenvalues coalesce. A circle
in the complex $\Gamma$ plane reveals the signature of four exchanging states
(see \fref{fig:ep4ab}(b)). Again the circle can be repeated in the complex
scattering plane (\fref{fig:ep4ab}(c)) resulting in the same exchange
behaviour.  In this case the question arises whether this is an EP4 or an EP5.
The exchange behaviour proves that the order of the exceptional point must be at
least four.  All wave functions of the participating states coalesce at the
critical point, however, to finally decide whether this is an EP5 further
perturbation parameter must be examined.

\begin{figure}
	\includegraphics[width=0.98\textwidth]{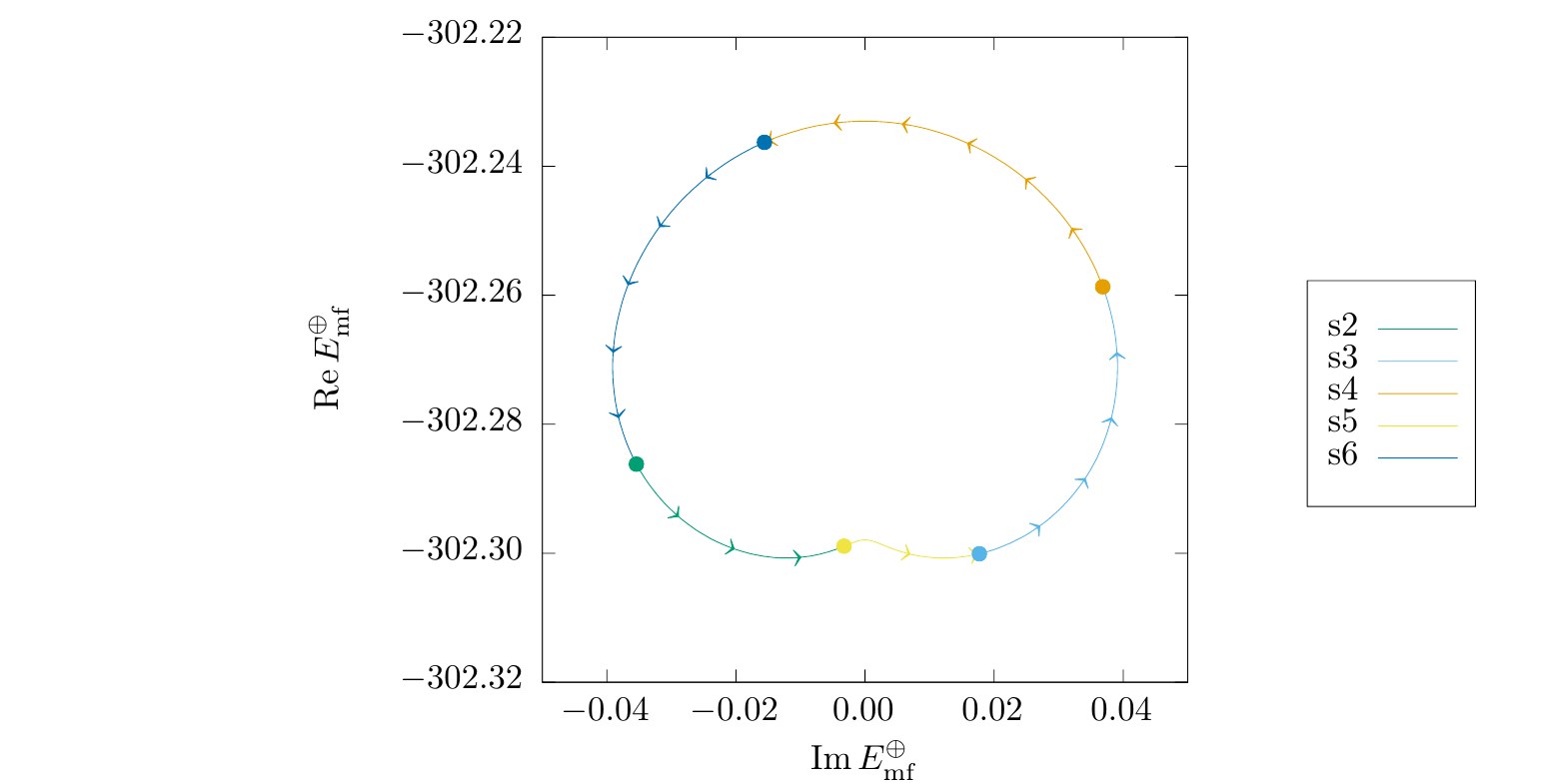}
	\caption{
		\label{fig:ep5sym}
		Mean-field energies $\bpparam{E_{\rm mf}}$ when the exceptional point where
		the bifurcations B2, B3a, and B3b coalesce is encircled in the complex
		symmetry-parameter space $s$ defined in \eref{eq:symparapath}.  }
\end{figure}

We introduce a new asymmetry parameter in \eref{eq:gpe}, which lifts or lowers
the potential wells
\begin{eqnarray}
	c_{{\rm g},1} = (V_0 + s) + \ii \Gamma
	~{\rm and}~
	c_{{\rm g},2} = (V_0 - s) + \ii \Gamma.
\end{eqnarray}
The new parameter $s$ allows us to break the remaining trap symmetry of the
system, e.g.\ one potential well is deepend, while the other is flattened. We
encircle the bifurcation for the scattering length $c_{\rm sc,crit,2} =
-0.93365$ and the coupling parameter $\Gamma = 1.03215$ on the path
\begin{eqnarray}
	\label{eq:symparapath}
	s = 5 \times 10^{-5} \ee^{\jj\phi} ~{\rm for}~
	\phi \in [0, 2\pi]
\end{eqnarray}
and observe a permutation of all five states with each other (see
\fref{fig:ep5sym}). Thus we have proven the existence of an exceptional
point of order five in this system.

\section{Summary and conclusion}

We have shown how the Gross-Pitaevskii equation for dipolar Bose-Einstein
condensates can be analytically continued with an ansatz of coupled Gaussians
using bicomplex numbers. Especially the representation in the idempotent basis
of the bicomplex numbers can be used to separate the bicomplex equations into
twice the number of coupled complex equations. This allows for the reuse of 
an algorithm developed for the integration of the complex equations.

By solving the analytically continued equations we have
shown that a dipolar Bose-Einstein condensate in a \PT-symmetric trap
exhibits a much richer bifurcation scenario than a
condensate without long-range interactions.  Most of the properties examined in
this paper have revealed interesting mathematical relations in the bicomplex
parameter space, which is experimentally inaccessible. However, the
understanding of the bifurcation scenrio is important since bifurcations
influence crucially the stabiliy of a condensate \cite{Haag14}.

Furthermore the additional states, while not experimentally accessible,
allow us to examine the order of the exceptional points.  Higher-order
exceptional points were found in other systems such as EP3s in the hydrogen
atom \cite{Holger2007} or in Bose-Einstein condensates \cite{Heiss13a}. Even
exeptional points of arbitrary order were shown to exist in a Bose-Hubbard
model \cite{Graefe08a}. However, no EPs with an order higher than three were
known for a BEC described by a mean-field model.

With the new (bicomplex) states we have demonstrated the properties of
the exceptional points associated with the bifurcations (\sref{sec:res}).
In particular the critical points where two or three bifurcation coalesce were
examined. By having shown that a parameter exists for which the
encircling of the critical value results in the permutation of all five states
participating in the exceptional point, we have indeed proven a fifth-order
exceptional point in dipolar Bose-Einstein condensates.

It was shown \cite{Brinker2015} that the signatures of exceptional points
influence the collapse dynamics of Bose-Einstein condensates. With the help of
a harmonic inversion analysis \cite{Fuchs2014} the order of degeneracies can be
translated into polynomial terms in the time-evolution. Further studies will
reveal the influence of such higher-order exceptional points on the collapse
dynamics.

% -----------------------------------------------------------------------------
\section*{References}

\bibliographystyle{unsrt}
\bibliography{paper}

\begin{thebibliography}{10}

\bibitem{Bender1999a}
C.~M. Bender, S.~Boettcher, and P.~N. Meisinger.
\newblock {$\mathcal{PT}$-symmetric quantum mechanics}.
\newblock {\em J. Math. Phys.}, 40:2201, 1999.

\bibitem{qm1}
H.~F. Jones and Jr. E.~S. Moreira.
\newblock {Q}uantum and classical statistical mechanics of a class of
  non-{H}ermitian {H}amiltonians.
\newblock {\em J. Phys. A}, 43:055307, 2010.

\bibitem{qm2}
V.~Jakubsk{\'y} and M.~Znojil.
\newblock {An explicitly solvable model of the spontaneous
  $\mathcal{PT}$-symmetry breaking}.
\newblock {\em Czech. J. Phys.}, 55:1113, 2005.

\bibitem{Mehri}
H.~Mehri-Dehnavi, A.~Mostafazadeh, and A.~Batal.
\newblock {Application of pseudo-{H}ermitian quantum mechanics to a complex
  scattering potential with point interactions}.
\newblock {\em J. Phys. A}, 43:145301, 2010.

\bibitem{Rueter10}
C.~E. Ruter, K.~G. Makris, R.~El-Ganainy, D.~N. Christodoulides, M.~Segev, and
  D.~Kip.
\newblock {Observation of parity-time symmetry in optics}.
\newblock {\em Nat. Phys.}, 6:192--195, 2010.

\bibitem{PhysRevA.88.053817}
I.~V. Barashenkov, G.~S. Jackson, and S.~Flach.
\newblock {Blow-up regimes in the $\mathcal{PT}$-symmetric coupler and the
  actively coupled dimer}.
\newblock {\em Phys. Rev. A}, 88:053817, 2013.

\bibitem{Deffner2015}
S.~Deffner and A.~Saxena.
\newblock Jarzynski equality in $\mathcal{P}\mathcal{T}$-symmetric quantum
  mechanics.
\newblock {\em Phys. Rev. Lett.}, 114:150601, 2015.

\bibitem{Albeverio2015}
S.~Albeverio, S.~Fassari, and F.~Rinaldi.
\newblock The discrete spectrum of the spinless one-dimensional {S}alpeter
  {H}amiltonian perturbed by $\delta$-interactions.
\newblock {\em J. Phys. A}, 48:185301, 2015.

\bibitem{Mostafazadeh2013b}
A.~Mostafazadeh.
\newblock Nonlinear spectral singularities for confined nonlinearities.
\newblock {\em Phys. Rev. Lett.}, 110:260402, 2013.

\bibitem{Bittner2012a}
S.~Bittner, B.~Dietz, U.~G{\"u}nther, H.~L. Harney, M.~Miski-Oglu, A.~Richter,
  and F.~Sch{\"a}fer.
\newblock {$\mathcal{PT}$ Symmetry and Spontaneous Symmetry Breaking in a
  Microwave Billiard}.
\newblock {\em Phys. Rev. Lett.}, 108:024101, 2012.

\bibitem{Schindler2011a}
J.~Schindler, A.~Li, M.~C. Zheng, F.~M. Ellis, and T.~Kottos.
\newblock {Experimental study of active $LRC$ circuits with $\mathcal{PT}$
  symmetries}.
\newblock {\em Phys. Rev. A}, 84:040101, 2011.

\bibitem{Schindler2012a}
J.~Schindler, Z.~Lin, J.~M. Lee, H.~Ramezani, F.~M. Ellis, and T.~Kottos.
\newblock {$\mathcal{PT}$-symmetric electronics}.
\newblock {\em J. Phys. A}, 45:444029, 2012.

\bibitem{0305-4470-38-9-L03}
A.~Ruschhaupt, F.~Delgado, and J.~G. Muga.
\newblock {Physical realization of $\mathcal{PT}$ -symmetric potential
  scattering in a planar slab waveguide}.
\newblock {\em J. Phys. A}, 38:L171, 2005.

\bibitem{opticPT1}
A.~Guo, G.~J. Salamo, D.~Duchesne, R.~Morandotti, M.~Volatier-Ravat, V.~Aimez,
  G.~A. Siviloglou, and D.~N. Christodoulides.
\newblock {Observation of $\mathcal{P}\mathcal{T}$-Symmetry Breaking in Complex
  Optical Potentials}.
\newblock {\em Phys. Rev. Lett.}, 103:093902, 2009.

\bibitem{ramezani10}
H.~Ramezani, T.~Kottos, R.~El-Ganainy, and D.~N. Christodoulides.
\newblock {Unidirectional nonlinear $\mathcal{PT}$-symmetric optical
  structures}.
\newblock {\em Phys. Rev. A}, 82:043803, 2010.

\bibitem{musslimani08a}
Z.~H. Musslimani, Konstantinos~G. Makris, R.~El-Ganainy, and D.~N.
  Christodoulides.
\newblock {Optical solitons in PT periodic potentials}.
\newblock {\em Phys. Rev. Lett.}, 100:30402, 2008.

\bibitem{optic1}
K.~G. Makris, R.~El-Ganainy, D.~N. Christodoulides, and Z.~H. Musslimani.
\newblock {$\mathcal{PT}$-Symmetric Periodic Optical Potentials}.
\newblock {\em Int. J. Theo. Phys.}, 50:1019, 2011.

\bibitem{optic2}
K.~G. Makris, R.~El-Ganainy, D.~N. Christodoulides, and Z.~H. Musslimani.
\newblock {Beam Dynamics in $\mathcal{PT}$ Symmetric Optical Lattices}.
\newblock {\em Phys. Rev. Lett.}, 100:103904, 2008.

\bibitem{Makris2010a}
K.~G. Makris, R.~El-Ganainy, D.~N. Christodoulides, and Z.H. Musslimani.
\newblock {$\mathcal{PT}$-symmetric optical lattices}.
\newblock {\em Phys. Rev. A}, 81:063807, 2010.

\bibitem{makris08}
K.~G. Makris, R.~El-Ganainy, D.~N. Christodoulides, and Z.H. Musslimani.
\newblock {Beam dynamics in PT symmetric optical lattices}.
\newblock {\em Phys. Rev. Lett.}, 100:103904, 2008.

\bibitem{Chong2011}
Y.~D. Chong, L.~Ge, and A.~D. Stone.
\newblock $\mathcal{P}\mathcal{T}$-symmetry breaking and laser-absorber modes
  in optical scattering systems.
\newblock {\em Phys. Rev. Lett.}, 106:093902, 2011.

\bibitem{Peng2014}
B.~Peng, S.~\"O. Kaya, F.~Lei, F.~Minufu, M.~Gianfreda, G.~L. Long, S.~Fan,
  F.~Nori, C.~M. Bender, and L.~Yang.
\newblock {Parity-time-symmetric whispering-gallery microcavities}.
\newblock {\em Nat. Phys.}, 10:394--398, 2014.

\bibitem{Dast2014}
D.~Dast, D.~Haag, H.~Cartarius, and G.~Wunner.
\newblock Quantum master equation with balanced gain and loss.
\newblock {\em Phys. Rev. A}, 90:052120, 2014.

\bibitem{Guthrlein2015}
R.~Gut\"ohrlein, J.~Schnabel, I.~Iskandarov, H.~Cartarius, J.~Main, and
  G.~Wunner.
\newblock {R}ealizing $\mathcal{PT}$-symmetric {BEC} subsystems in closed
  {H}ermitian systems.
\newblock {\em Journal of Physics A: Mathematical and Theoretical}, 48:335302,
  2015.

\bibitem{And95a}
M.~H. Anderson, J.~R. Ensher, M.~R. Matthews, C.~E. Wieman, and E.~A. Cornell.
\newblock Observation of {B}ose-{E}instein condensation in a dilute atomic
  vapor.
\newblock {\em Science}, 269:198, 1995.

\bibitem{Bra95a}
C.~C. Bradley, C.~A. Sackett, J.~J. Tollett, and R.~G. Hulet.
\newblock Evidence of {B}ose-{E}instein condensation in an atomic gas with
  attractive interactions.
\newblock {\em Phys. Rev. Lett.}, 75:1687, 1995.

\bibitem{Dav95a}
K.~B. Davis, M.~O. Mewes, M.~R. Andrews, N.~J. van Druten, D.~S. Durfee, D.~M.
  Kurn, and W.~Ketterle.
\newblock Bose-{E}instein condensation in a gas of sodium atoms.
\newblock {\em Phys. Rev. Lett.}, 75:3969, 1995.

\bibitem{Griesmaier2005}
A.~Griesmaier, J.~Werner, S.~Hensler, J.~Stuhler, and T.~Pfau.
\newblock {B}ose-{E}instein {C}ondensation of {C}hromium.
\newblock {\em Phys. Rev. Lett.}, 94:160401, 2005.

\bibitem{Santos2000}
L.~Santos, G.~V. Shlyapnikov, P.~Zoller, and M.~Lewenstein.
\newblock {B}ose-{E}instein {C}ondensation in {T}rapped {D}ipolar {G}ases.
\newblock {\em Phys. Rev. Lett.}, 85:1791--1794, 2000.

\bibitem{PhysRevA.77.061601}
Q.~Beaufils, R.~Chicireanu, T.~Zanon, B.~Laburthe-Tolra, E.~Mar\'echal,
  L.~Vernac, J.-C. Keller, and O.~Gorceix.
\newblock {A}ll-optical production of chromium {B}ose-{E}instein condensates.
\newblock {\em Phys. Rev. A}, 77:061601, 2008.

\bibitem{PhysRevLett.107.190401}
M.~Lu, N.~Q. Burdick, S.~H. Youn, and B.~L. Lev.
\newblock Strongly dipolar bose-einstein condensate of dysprosium.
\newblock {\em Phys. Rev. Lett.}, 107:190401, 2011.

\bibitem{PhysRevLett.101.080401}
T.~Lahaye, J.~Metz, B.~Fr\"ohlich, T.~Koch, M.~Meister, A.~Griesmaier, T.~Pfau,
  H.~Saito, Y.~Kawaguchi, and M.~Ueda.
\newblock $d$-{W}ave {C}ollapse and {E}xplosion of a {D}ipolar
  {B}ose-{E}instein {C}ondensate.
\newblock {\em Phys. Rev. Lett.}, 101:080401, 2008.

\bibitem{Raghavan1999}
S.~Raghavan, A.~Smerzi, S.~Fantoni, and S.~R. Shenoy.
\newblock {C}oherent oscillations between two weakly coupled {B}ose-{E}instein
  condensates: {J}osephson effects, $\ensuremath{\pi}$ oscillations, and
  macroscopic quantum self-trapping.
\newblock {\em Phys. Rev. A}, 59:620--633, 1999.

\bibitem{Zaman2009}
M.~Asad-uz Zaman and D.~Blume.
\newblock {A}ligned dipolar {B}ose-{E}instein condensate in a double-well
  potential: {F}rom cigar shaped to pancake shaped.
\newblock {\em Phys. Rev. A}, 80:053622, 2009.

\bibitem{Xiong2009}
B.~Xiong, J.~Gong, H.~Pu, W.~Bao, and B.~Li.
\newblock {S}ymmetry breaking and self-trapping of a dipolar {B}ose-{E}instein
  condensate in a double-well potential.
\newblock {\em Phys. Rev. A}, 79:013626, 2009.

\bibitem{Gericke2008}
T.~Gericke, P.~W\"urtz, D.~Reitz, T.~Langen, and H.~Ott.
\newblock {H}igh-resolution scanning electron microscopy of an ultracold
  quantum gas.
\newblock {\em Nat. Phys.}, 4:949--953, 2008.

\bibitem{Robins2008}
N.~P. Robins, C.~Figl, M.~Jeppesen, and G.R. Dennis.
\newblock {A} pumped atom laser.
\newblock {\em Nat. Phys.}, 4:731--736, 2008.

\bibitem{Schneble2004}
D.~Schneble, G.~K. Campbell, E.~W. Streed, M.~Boyd, D.~E. Pritchard, and
  W.~Ketterle.
\newblock Raman amplification of matter waves.
\newblock {\em Phys. Rev. A}, 69:041601, 2004.

\bibitem{Yoshikawa2004}
Y.~Yoshikawa, T.~Sugiura, Y.~Torii, and T.~Kuga.
\newblock {O}bservation of superradiant {R}aman scattering in a
  {B}ose-{E}instein condensate.
\newblock {\em Phys. Rev. A}, 69:041603, 2004.

\bibitem{Kreibich2014}
M.~Kreibich, J.~Main, H.~Cartarius, and G.~Wunner.
\newblock {Realizing $\mathcal{PT}$-symmetric non-Hermiticity with ultracold
  atoms and Hermitian multiwell potentials}.
\newblock {\em Phys. Rev. A}, 90:033630, 2014.

\bibitem{PhysRevLett.98.030406}
S.~Ronen, D.~C.~E. Bortolotti, and J.~L. Bohn.
\newblock Radial and angular rotons in trapped dipolar gases.
\newblock {\em Phys. Rev. Lett.}, 98:030406, 2007.

\bibitem{Fortanier2014}
R.~Fortanier, D.~Dast, D.~Haag, H.~Cartarius, J.~Main, G.~Wunner, and
  R.~Gut\"ohrlein.
\newblock {D}ipolar {B}ose-{E}instein condensates in a $\mathcal{PT}$-symmetric
  double-well potential.
\newblock {\em Phys. Rev. A}, 89:063608, 2014.

\bibitem{Kato66}
T.~Kato.
\newblock {\em Perturbation theory for linear operators}.
\newblock Springer, Berlin, 1966.

\bibitem{Holger2007}
H.~Cartarius, J.~Main, and G.~Wunner.
\newblock {E}xceptional {P}oints in {A}tomic {S}pectra.
\newblock {\em Phys. Rev. Lett.}, 99:173003, 2007.

\bibitem{Heiss2008}
W.~D. Heiss.
\newblock Chirality of wavefunctions for three coalescing levels.
\newblock {\em Journal of Physics A: Mathematical and Theoretical}, 41:244010,
  2008.

\bibitem{Graefe08a}
E.~M. Graefe, U.~G{\"u}nther, H.~J. Korsch, and A.~E. Niederle.
\newblock {A non-{H}ermitian $\mathcal{PT}$ symmetric {B}ose-{H}ubbard model:
  eigenvalue rings from unfolding higher-order exceptional points}.
\newblock {\em J. Phys. A}, 41:255206, 2008.

\bibitem{Dem12}
G.~Demange and E.-M. Graefe.
\newblock {S}ignatures of three coalescing eigenfunctions.
\newblock {\em J. Phys. A: Math. Theor.}, 45:025303, 2012.

\bibitem{Heiss2012}
W.~D. Heiss.
\newblock The physics of exceptional points.
\newblock {\em Journal of Physics A: Mathematical and Theoretical}, 45:444016,
  2012.

\bibitem{Heiss13a}
W.~D. Heiss, H.~Cartarius, G.~Wunner, and J.~Main.
\newblock {Spectral singularities in {${\mathcal P}{\mathcal T}$}-symmetric
  Bose--Einstein condensates}.
\newblock {\em J. Phys. A}, 46:275307, 2013.

\bibitem{Guthrlein2013}
R.~Gut\"ohrlein, J.~Main, C.~Cartarius, and G.~Wunner.
\newblock Bifurcations and exceptional points in dipolar {B}ose-{E}instein
  condensates.
\newblock {\em Journal of Physics A: Mathematical and Theoretical}, 46:305001,
  2013.

\bibitem{Lahaye2009}
T.~Lahaye, C.~Menotti, L.~Santos, M.~Lewenstein, and T.~Pfau.
\newblock The physics of dipolar bosonic quantum gases.
\newblock {\em Reports on Progress in Physics}, 72:126401, 2009.

\bibitem{McLachlan1964a}
A.~D. McLachlan.
\newblock {A variational solution of the time-dependent Schrodinger equation}.
\newblock {\em Mol. Phys.}, 8:39--44, 1964.

\bibitem{Rau10c}
S.~Rau, J.~Main, P.~K\"oberle, and G.~Wunner.
\newblock {P}itchfork bifurcations in blood-cell-shaped dipolar
  {B}ose-{E}instein condensates.
\newblock {\em Phys. Rev. A}, 81:031605, 2010.

\bibitem{Dast2013}
D.~Dast, D.~Haag, H.~Cartarius, J.~Main, and G.~Wunner.
\newblock {E}igenvalue structure of a {B}ose-{E}instein condensate in a
  {$\mathcal{PT}$} -symmetric double well.
\newblock {\em J. Phys. A}, 46:375301, 2013.

\bibitem{LUNA-ELIZARRARAS2012}
M.E. Shapiro, D.C. Struppa, and A.~Vajiac.
\newblock {Bicomplex Numbers and their Elementary Functions}.
\newblock {\em {Cubo (Temuco)}}, 14:61 -- 80, 2012.

\bibitem{Brody2011a}
D.~C. Brody and E.-M. Graefe.
\newblock On complexified mechanics and coquaternions.
\newblock {\em Journal of Physics A: Mathematical and Theoretical}, 44:072001,
  2011.

\bibitem{brody2011b}
D.~C. Brody and E.-M. Graefe.
\newblock Coquaternionic quantum dynamics for two-level systems.
\newblock {\em Acta Polytechnica}, 51, 2011.

\bibitem{Bagchi2015}
B.~Bagchi and A.~Banerjee.
\newblock Bicomplex hamiltonian systems in quantum mechanics.
\newblock {\em Journal of Physics A: Mathematical and Theoretical}, 48:505201,
  2015.

\bibitem{FortanierDoc}
R.~M. Fortanier.
\newblock {\em {V}ariational approaches to dipolar {B}ose-{E}instein
  condensates}.
\newblock PhD thesis, Universit\"at Stuttgart, 2014.
\newblock http://elib.uni-stuttgart.de/opus/volltexte/2014/9223.

\bibitem{Carlson}
{B. C. Carlson}.
\newblock {N}umerical computation of real or complex elliptic integrals.
\newblock {\em Numerical Algorithms}, 10:13, 1995.

\bibitem{Graefe12}
E.-M. Graefe.
\newblock {Stationary states of a $\mathcal{PT}$ symmetric two-mode
  Bose--Einstein condensate}.
\newblock {\em J. Phys. A}, 45:444015, 2012.

\bibitem{Cartarius12b}
H.~Cartarius and G.~Wunner.
\newblock {Model of a $\mathcal{PT}$-symmetric Bose-Einstein condensate in a
  $\delta$-function double-well potential}.
\newblock {\em Phys. Rev. A}, 86:013612, 2012.

\bibitem{Dast13}
D.~Dast, D.~Haag, H.~Cartarius, G{\"u}nter Wunner, R.~Eichler, and J.~Main.
\newblock {A Bose-Einstein condensate in a $\mathcal{PT}$-symmetric double
  well}.
\newblock {\em Fortschritte der Physik}, 61:124--139, 2013.

\bibitem{Haag14}
D.~Haag, D.~Dast, A.~L{\"o}hle, H.~Cartarius, J.~Main, and G.~Wunner.
\newblock {Nonlinear quantum dynamics in a $\mathcal{PT}$-symmetric double
  well}.
\newblock {\em Phys. Rev. A}, 89:023601, 2014.

\bibitem{Brinker2015}
J.~Brinker, J.~Fuchs, J.~Main, G.~Wunner, and H.~Cartarius.
\newblock {V}erification of exceptional points in the collapse dynamics of
  {B}ose-{E}instein condensates.
\newblock {\em Phys. Rev. A}, 91:013609, 2015.

\bibitem{Fuchs2014}
J.~Fuchs, J.~Main, H.~Cartarius, and G.~Wunner.
\newblock {H}armonic inversion analysis of exceptional points in resonance
  spectra.
\newblock {\em Journal of Physics A: Mathematical and Theoretical}, 47:125304,
  2014.

\end{thebibliography}

\end{document}